\documentclass[a4paper]{jpconf}

\usepackage{graphicx}

\begin{document}

\title{Dynamical reduction models: present status and future developments}

\author{Angelo Bassi}

\address{Dipartimento di Fisica Teorica,
Universit\`a degli Studi di Trieste, Strada Costiera 11, 34014
Trieste, Italy. \\  Mathematisches Institut der Ludwig-Maximilians
Universit\"at, Theresienstr. 39, 80333 M\"unchen, Germany.}

\ead{bassi@ts.infn.it, bassi@mathematik.uni-muenchen.de}

\begin{abstract}
We review the major achievements of the dynamical reduction program,
showing why and how it provides a unified, consistent description of
physical phenomena, from the microscopic quantum domain to the
macroscopic classical one. We discuss the difficulties in
generalizing the existing models in order to comprise also
relativistic quantum field theories. We point out possible future
lines of research, ranging from mathematical physics to
phenomenology.
\end{abstract}

\section{Quantum Mechanics, measurements and environment}

Standard Quantum Mechanics is known to talk only about the
outcomes of measurements, but it has nothing to say about the
world as it is, independently of any measurement or act of
observation. This is a source of serious difficulties, which have
been clearly elucidated e.g. by J. Bell \cite{bell1}: {\it It
would seem that the theory is exclusively concerned about `results
of measurements', and has nothing to say about anything else. What
exactly qualifies some physical systems to play the role of
`measurer'? Was the wavefunction  of the world waiting to jump for
thousands of millions of years until a single-celled living
creature appeared? Or did it have to wait a little bit longer, for
some better qualified system ... with a Ph.D.?}

Measuring devices, like photographic plates and bubble chambers, are
very sophisticated and highly structured physical systems, which
anyhow are made up of atoms; we then expect them to be ultimately
described in quantum mechanical terms by means of the Schr\"odinger
equation. What else should we do, taking into account that people
are trying to describe also the entire universe quantum
mechanically? But if we describe measurements in this way, then we
do {\it not} get any outcome at the end of the process. The
Schr\"odinger equation is linear, the superposition principle enters
into play and it does it in such a way that all possible outcomes
are there simultaneously in the wave function, but none of them is
selected as the one which occurs physically. Yet, if we perform a
measurement, we always get a definite outcome. So we have a problem
with Quantum Mechanics.

In recent years, a not so new idea is gaining more and more
credit: measuring devices are different from microscopic systems:
they are big objects which unavoidably interact with the
surrounding environment. Such an interaction turns out to be very
peculiar because it destroys the coherence between different terms
of a superposition and seems to reduce a pure state, where all
terms of the superposition are there simultaneously, into a
statistical mixture of the states, and moreover it does so with
the correct quantum probabilities. What else do we need?

This idea, that the environment somehow {\it naturally} guarantees
the emergence of definite properties when moving from the micro to the
macro, by destroying coherence among different terms of a
superposition, is very appealing. But wrong. I will not spend much
time on this issue, since many papers already have appeared on the
subject, starting from those of Bell~\cite{bell2} to very recent
ones~\cite{bg1,leg,ad1}. I note here that the division between a
system and its environment is not a division dictated by Nature.
Such a division is arbitrarily set by the Physicist because he or
she is not able to solve the Schr\"odinger equation for the global
system; he or she then decides to select some degrees of freedom as
the relevant ones, and to trace over all other degrees. This is a
very legitimate division, but not compelling at all. Such a division
is more or less equivalent to the division between a quantum system
and a measuring device: it's artificial, just a matter of practical
convenience. But if the physicist were able to analyze exactly the
microscopic quantum system, the macroscopic apparatus {\it and} the
surrounding environment together, i.e. if he or she used the
Schr\"odinger equation to study the global system, he or she would
get a very simple result: once more, because of linearity, all terms
of the superposition would be present at the same time in the wave
function, no one of them being singled out as that which really
occurs when the measurement is performed in the laboratory.

The so called {\it measurement problem} of Quantum Mechanics is an
open problem still waiting for a solution. Dynamical reduction
models, together with Bohmian Mechanics, up to now are, in my
opinion, the most serious candidates for a resolution of this
problem.

\section{The dynamical reduction program}

Continuing quoting Bell: {\it If the theory is to apply to anything
but highly idealized laboratory operations, are we not obliged to
admit that more or less `measurement-like' processes are going on
more or less all the time, more or less everywhere? Do we not have
jumping then all the time?} The basic idea behind the dynamical
reduction programm is precisely this: spontaneous and random
collapses of the wave function occur all the time, to all particles,
whether isolated or interacting, whether they form just a tiny atom
or a big measuring device. Of course, such collapses must be rare
and mild for microscopic systems, in order not to disturb their
quantum behavior as predicted by the Schr\"odinger equation. At the
same time, their effect must add up in such a way that, when
thousands of millions of particles are glued together to form a
macroscopic system, a single collapse occurring to one of the
particles affects the global system. We then have thousands of
millions of such collapses acting very frequently on the
macro-system, which together force its wave function to be very well
localized in space.

The aim of the dynamical reduction programm is then to modify the
Schr\"odinger evolution, by introducing new terms having the
following properties:
\begin{itemize}
\item They must be {\it non-linear}, as one wants to break the
superposition principle at the macroscopic level and assure the
localization of the wave function of macro-objects.
\item They must be {\it stochastic} because, when describing
measurement-like situations, one needs to explain why the outcomes
occur randomly; more than this, one needs to explain why they are
distributed according to the Born probability rule.
\item There must be an {\it amplification mechanism} according to
which the new terms have negligible effects on the dynamics of
microscopic systems but, at the same time, their effect becomes
very strong for large many-particle systems such as macroscopic
objects, in order to recover their classical-like behavior.
\end{itemize}
If we look carefully at these requirement, we realize that they are
very mandatory: there is no assurance at all beforehand, that they
can be consistently fulfilled. I think that one of the greatest
merits of the GRW proposal \cite{grw} is to have shown that they can
be implemented in a consistent and satisfactory model.

\section{The GRW model}

Let us consider a system of $N$ particles which, only for
simplicity's sake, we take to be scalar; the GRW model is defined
by the following postulates:
\begin{description}
\item[States.] The state of the system is represented by a wave
function $\psi({\bf x}_{1}, {\bf x}_{2}, \ldots {\bf x}_{N})$
belonging to the Hilbert space ${\mathcal L}^2({\bf R}^{3N})$.
\item[Evolution.] At  random times, each particle experiences a sudden jump of
the form:
\begin{equation}
\psi_{t}({\bf x}_{1}, {\bf x}_{2}, \ldots {\bf x}_{N}) \quad
\longrightarrow \quad \frac{L_{n}({\bf x}) \psi_{t}({\bf x}_{1},
{\bf x}_{2}, \ldots {\bf x}_{N})}{\|L_{n}({\bf x}) \psi_{t}({\bf
x}_{1}, {\bf x}_{2}, \ldots {\bf x}_{N})\|},
\end{equation}
where $\psi_{t}({\bf x}_{1}, {\bf x}_{2}, \ldots {\bf x}_{N})$ is
the statevector of the whole system at time $t$, immediately prior
to the jump process. $L_{n}({\bf x})$ is a linear operator which is
conventionally chosen equal to:
\begin{equation}
L_{n}({\bf x}) \quad = \quad \sqrt[4]{\left(
\frac{\alpha}{\pi}\right)^3} \exp \left[ - \frac{\alpha}{2} ({\bf
q}_{n} - {\bf x})^2 \right],
\end{equation}
where $\alpha$ is a new parameter of the model which sets the the
width of the localization process, and ${\bf q}_{n}$ is the position
operator associated to the $n$-th particle; the random variable
${\bf x}$ corresponds to the place where the jump occurs. Between
two consecutive jumps, the statevector evolves according to the
standard Schr\"odinger equation.

The probability density for a jump taking place at the position
${\bf x}$ for the $n$-th particle is given by:
\begin{equation}
p_{n}({\bf x}) \quad \equiv \quad \|L_{n}({\bf x}) \psi_{t}({\bf
x}_{1}, {\bf x}_{2}, \ldots {\bf x}_{N})\|^2,
\end{equation}
and the probability densities for the different particles are
independent.

Finally, it is assumed that the jumps are distributed in time like
a Poissonian process with frequency $\lambda$, which is the second
new parameter of the model.

The standard numerical values for $\alpha$ and $\lambda$ are:
\begin{equation} \label{eq:num}
\lambda \; \simeq \; 10^{-16} \, \makebox{sec$^{-1}$}, \quad\qquad
\alpha \; \simeq \; 10^{10} \, \makebox{cm$^{-2}$}.
\end{equation}

\item[Ontology.] Let the $m_{n}$ be the mass associated to
the $n$-th ``particle'' of the system (I should say: to what is
called ``a particle'', according to the standard terminology); then
the function:
\begin{equation}
\rho^{(n)}_{t}({\bf x}_{n}) \; \equiv \; m_{n} \int d^3 x_{1} \ldots
d^3 x_{n-1} d^2 x_{n+1} \ldots d^3 x_{N} \, | \psi_{t}({\bf x}_{1},
{\bf x}_{2}, \ldots {\bf x}_{N}) |^2
\end{equation}
represents the {\it density of mass} \cite{int} of that ``particle''
in space, at time $t$.
\end{description}
These are the axioms of the GRW model: as we see, words such as
`measurement', `observation', `macroscopic', `environment' do not
appear. There is only a {\it universal} dynamics governing all
physical processes, and an ontology which tells how the physical
world is, according to the model, independently of any act of
observation.

The GRW model, together with other dynamical reduction models
which have appeared in the literature, has been extensively
studied (see \cite{rev1} and \cite{rev2} for a review on this
topic); in particular---with the numerical choice for $\lambda$
and $\alpha$ given in~(\ref{eq:num})---the following three
important properties have been proved, which we will state in more
quantitative terms in the following section:
\begin{itemize}
\item At the microscopic level, quantum systems behave almost
exactly as predicted by standard Quantum Mechanics, the
differences being so tiny that they can hardly be detected with
present-day technology.
\item At the macroscopic level, wave functions of macro-objects
are almost always very well localized in space, so well localized
that their centers of mass behave, for all practical purposes,
like point-particles moving according to Newton's laws.
\item In a measurement-like situation, e.g. of the von Neumann type,
GRW reproduces---as a consequence of the modified dynamics---both
the Born probability rule and the postulate of wave-packet
reduction.
\end{itemize}
Accordingly, models of spontaneous wave function collapse provide
a unified description of all physical phenomena, at least at the
non-relativistic level, and a consistent solution to the
measurement problem of Quantum Mechanics.

It may be helpful to stress some points about the world-view
provided by the GRW model. According to the interpretation given by
the third axiom, there are no particles at all in the theory! There
are only distributions of masses which, at the microscopic level,
are in general quite spread out. An electron, for example, is not a
point following a trajectory---as it would be in Bohmian
Mechanics---but a wavy system diffusing in space. When, in a
double-slit experiment, it goes through the apertures, it literarily
goes through both of them, as a classical water-wave would do. The
peculiarity of the electron, which qualifies it as a quantum system,
is that when we try to localize it in space by letting it
interacting with a measuring device, e.g. a photographic plate,
then, according to the second axiom and because of the interaction
with the plate, its wave function very rapidly shrinks in space till
is gets localized to a spot, the spot where the plate is impressed
and which represents the outcome of the measurement. Such a behavior
is not postulated {\it ad hoc} as done in standard Quantum
Mechanics; it is a direct consequence of the universal dynamics of
the GRW model.

Also macroscopic objects are waves; their centers of mass are not
mathematical points, rather they are represented by some function
defined throughout space. But macro-objects have a nice property:
according to the GRW dynamics, each of them is always almost
perfectly located in space, which means that the wave functions
associated to their centers of mass are appreciably different from
zero only within a very tiny region of space (whose linear extension
is of order $10^{-14}$ m or smaller, as we shall see), so tiny that
they can be considered point-like for all practical purposes. This
is the reason why Newton's mechanics of point particles is such a
satisfactory theory for macroscopic classical systems.

Even though the GRW model contains no particles at all, we will
keep referring to micro-system as `particles', just for a matter
of convenience.

\section{Dynamical reduction models and stochastic differential equations}

The second axiom of the GRW model concerning the evolution of
physical systems can be written more succinctly in terms of
stochastic differential equations. According to the QMUPL model
first proposed in \cite{di1} and subsequently studied in
\cite{ab1} (see also references therein), a wave function
$\psi_{t}(\{ x \}) \equiv \psi_{t}(x_{1}, x_{2}, \ldots x_{N})$
evolves according to the following stochastic differential
equation, where for simplicity we assume the dynamics to take
place only in one dimension:
\begin{equation} \label{nlemp}
d\,\psi_{t}(\{ x \}) \; = \; \left[ -\frac{i}{\hbar}\,
H_{\makebox{\tiny TOT}}\, dt + \sum_{n=1}^{N}\sqrt{\lambda_{n}}\,
( q_{n} - \langle q_{n} \rangle_{t})\, d W_{t}^{n} -
\frac{1}{2}\,\sum_{n=1}^{N} \lambda_{n} ( q_{n} - \langle q_{n}
\rangle_{t})^2 dt \right] \psi_{t}(\{ x \});
\end{equation}
$H_{\makebox{\tiny TOT}}$ is the standard quantum Hamiltonian of
the composite system; the symbol $\langle q_{n} \rangle_{t}$
represents the quantum average $\langle \psi_{t} | q_{n} |
\psi_{t} \rangle$ of the position operator $q_{n}$; the random
processes $W_{t}^{n}$ ($n = 1, \ldots N$) are $N$ independent
standard Wiener processes defined on a probability space $(\Omega,
{\mathcal F}, {\bf P})$, and the coupling constants $\lambda_{n}$
are defined as follows:
\begin{equation} \label{eq:lam}
\lambda_{n} \; \equiv \; \frac{m_{n}}{m_{0}}\, \lambda_{0},
\end{equation}
where $m_{n}$ is the mass of the $n$-th particle, while $m_{0}$ is a
reference mass which we assume equal to the mass of a nucleon:
$m_{0} \simeq 1.7 \times 10^{-27}$ Kg. In order for the QMUPL model
to be empirically equivalent to the GRW model, one has to choose
$\lambda_{0} \simeq 10^{-2}$ m$^{-2}$ sec$^{-1}$.

The above equation has been studied quite in detail in the
literature; the behavior of microscopic systems and macroscopic
objects, and in particular of measurement-like situations, is the
following.

\subsection{Microscopic systems.}

According to the dynamical reduction program, microscopic quantum
systems have an existence on their own, independently of any act
of observation. Anyway, they cannot be seen directly, and in order
to discover their properties they have to be subjected to
measurements.

As shown in \cite{mis}, measurable quantities are given by averages
of the form ${\bf E}[\langle O \rangle_{t}]$, where $O$ is (in
principle) any self--adjoint operator and ${\bf E}[\ldots]$ denotes
the stochastic average. It is not difficult to prove that ${\bf
E}[\langle O \rangle_{t}] =\makebox{Tr} [ O \rho_{t} ]$ where the
statistical operator $\rho_{t} \equiv {\bf E} [
|\psi_{t}\rangle\langle\psi_{t}| ]$ satisfies the Lindblad--type
equation:
\begin{equation} \label{lin}
\frac{d}{dt}\, \rho_{t} \; = \; -\frac{i}{\hbar}\, \left[ H,
\rho_{t} \right] - \frac{1}{2} \sum_{n=1}^{N} \lambda_{n} \left[
q_{n}, \left[ q_{n}, \rho_{t} \right] \right].
\end{equation}

This is the master equation first introduced by Joos and
Zeh~\cite{dec2} to describe the interaction between quantum
particles with a surrounding environment; consequently, only as
far as experimental results are concerned, the model behaves as if
the system were an open quantum system, even though in our case an
environment need {\it not} be present for the collapses to occur.
Given this, an easy way to understand the magnitude of the
physical effects of the reduction process is to compare the
strength of the collapse mechanism (measured by the constants
$\lambda_{n}$) with the loss of coherence due to the presence of
an environment.

Such a comparison is given in Table \ref{tab1}, when the system
under study is a very small particle like an electron, or an almost
macroscopic object like a dust particle.
\begin{table}
\caption{\label{tab1} Decoherence rates (in
cm${}^{-2}$sec${}^{-1}$) for different kinds of scattering
processes (taken from Joos and Zeh \cite{dec2}). In the last line:
$\lambda_{n}$ (in cm${}^{-2}$sec${}^{-1}$) as defined in
(\ref{eq:lam}).}
\begin{center}
\begin{tabular}{lcc} \br
Cause of decoherence & $\quad$ $10^{-3}$ cm  $\quad$ & $\quad$ $10^{-6}$ cm $\quad$ \\
 & dust particle  & large molecule \\ \mr
Air molecules  & $10^{36}$ & $10^{30}$  \\
Laboratory vacuum  & $10^{23}$ & $10^{17}$ \\
Sunlight on earth  & $10^{21}$ & $10^{13}$ \\
300K photons  & $10^{19}$ & $10^{6}$ \\
Cosmic background rad. & $10^{6}$ & $10^{-12}$ \\
\mr
COLLAPSE  & $10^{7}$ & $10^{-2}$ \\
\br
\end{tabular}
\end{center}
\end{table}
We see that, for most sources of decoherence, the experimentally
testable effects of the collapse mechanism are weaker than the
disturbances produced by the interaction of the system with a
surrounding environment. This implies that, in order to test the
GRW effects, one has to keep a quantum system isolated for a
sufficiently long time, from most sources of decoherence, and this
is difficult to achieve, unless very sophisticated experiments are
performed (more about this in the following). The analysis then
shows that the predictions of the GRW model are in good agreement
with standard quantum mechanical predictions.

\subsection{Macroscopic objects}

Let us now consider what happens not to a small quantum system,
but to a macroscopic object. For the purposes of our analysis, it
is convenient to switch to the center--of--mass ($R$) and relative
($\tilde{x}_{1}, \tilde{x}_{2}, \ldots \tilde{x}_{N}$)
coordinates:
\begin{equation}
R \; = \; \frac{1}{M}\, \sum_{n=1}^{N} m_{n}\, x_{n} \qquad x_{n}
\; = \; R + \tilde{x}_{n}, \qquad\quad M \; = \; \sum_{n=1}^{N}
m_{n};
\end{equation}
let $Q$ be the position operator for the center of mass and
$\tilde{q} _{n}$ ($n = 1 \ldots N$) the position operators
associated to the relative coordinates. It is not difficult to
show that, under the assumption $H_{\makebox{\tiny TOT}} =
H_{\makebox{\tiny CM}} + H_{\makebox{\tiny rel}}$, the dynamics
for the center of mass and that for the relative motion decouple;
in other words, $\psi_{t}(\{ x \}) = \psi_{t}^{\makebox{\tiny
CM}}(R) \otimes \psi_{t}^{\makebox{\tiny rel}}(\{\tilde{x}\})$
solves Eq. (\ref{nlemp}) whenever $\psi_{t}^{\makebox{\tiny
CM}}(R)$ and $\psi_{t}^{\makebox{\tiny rel}}(\{\tilde{x}\})$
satisfy the following equations:
\begin{eqnarray}
d\psi_{t}^{\makebox{\tiny rel}}(\{\tilde{x}\}) & = & \left[
-\frac{i}{\hbar}\, H_{\makebox{\tiny rel}}\, dt +
\sum_{n=1}^{N}\sqrt{\lambda_{n}}\, ({\tilde{q}}_{n} - \langle
{\tilde{q}}_{n} \rangle_{t}) d W_{t}^{n} -
\frac{1}{2}\,\sum_{n=1}^{N} \lambda_{n} ({\tilde{q}}_{n} - \langle
{\tilde{q}}_{n} \rangle_{t})^2 dt
\right] \psi_{t}^{\makebox{\tiny rel}}(\{\tilde{x}\}), \nonumber \\ & & \\
d\psi_{t}^{\makebox{\tiny CM}}(R) & = & \left[ -\frac{i}{\hbar}\,
H_{\makebox{\tiny CM}}\, dt + \sqrt{\lambda_{\makebox{\tiny
CM}}}\, (Q - \langle Q \rangle_{t}) d W_{t} -
\frac{\lambda_{\makebox{\tiny CM}}}{2}\, (Q - \langle Q
\rangle_{t})^2 dt \right] \psi_{t}^{\makebox{\tiny CM}}(R),
\label{eqcm}
\end{eqnarray}
with:
\begin{equation}
\lambda_{\makebox{\tiny CM}} \; = \; \sum_{n=1}^{N} \lambda_{n} \; =
\; \frac{M}{m_{0}}\, \lambda_{0}.
\end{equation}
The first of the above equations describes the internal motion of
the system: it basically tells that, since the constants
$\lambda_{n}$ are very small in magnitude, the internal structure is
described in agreement with the standard Schr\"odinger equation,
modulo small deviations of the type discussed in the previous
subsection. We now focus our attention on the second equation.

Eq.~(\ref{eqcm}) shows that the reducing terms associated to the
center of mass of a composite system are equal to those associated
to a particle having mass equal to the total mass $M$ of the whole
system. The constant $\lambda_{\makebox{\tiny CM}}$ has now a much
larger value than that of the $\lambda_{n}$, thus we expect the
dynamics of the center of mass to be completely different from that
of the microscopic quantum particles discussed in the previous
section. This is precisely the {\it amplification mechanism} we
talked about before: tiny collapses associated to each particle sum
up and produce a very strong collapse on the global system.

As a matter of fact, in Ref.~\cite{ab1} it has been proven that,
for macroscopic values of $M$, an initially spread wave function
is very rapidly localized in space, within a time interval much
smaller than the perception time of a human observer, and it
reaches asymptotically the value (for an isolated system)
\begin{equation} \label{aval1}
\sigma_{q}(\infty) \quad \simeq \quad \left( 1.5 \times 10^{-15}
\sqrt{\frac{\makebox{Kg}}{M}}\, \right)\, \makebox{m} \quad \simeq
\quad \left\{
\begin{array}{ll}
4.6 \times 10^{-14} &
\makebox{m $\quad$ for an 1--g object,} \\
& \\ 5.9 \times 10^{-28} & \makebox{m $\quad$ for the Earth.}
\end{array}
\right.
\end{equation}
As we see, the asymptotic spread of the wave function of the
center of mass of a macroscopic object is very very small, so
small that the wave function can be considered, for all practical
purposes like a point in space! A similar localization occurs also
in momentum space, within the limits allowed by Hesienberg's
uncertainty principle. This is how dynamical reduction models
justify the point-like behavior of macroscopic classical
particles.

But particles move in space: do they move according to Newton's
laws? It is easy to show that the average value ${\bf E}[\langle Q
\rangle_{t}]$ of the mean position and ${\bf E}[\langle P
\rangle_{t}]$ of the mean momentum satisfy the following
equations:
\begin{equation}
\frac{d}{dt}\,{\bf  E}\left[ \langle Q \rangle_{t} \right] \; = \;
\frac{i}{\hbar} {\bf  E}\left[ \langle [H_{\makebox{\tiny CM}}, Q]
\rangle_{t} \right], \qquad\qquad \frac{d}{dt}\,{\bf E}\left[
\langle P \rangle_{t} \right] \; = \; \frac{i}{\hbar} {\bf  E}\left[
\langle [H_{\makebox{\tiny CM}}, P] \rangle_{t} \right],
\label{emim}
\end{equation}
which can be considered as the stochastic extension of Ehrenfest's
theorem; we then recover the classical equation of motion, in the
appropriate limit. But this is not enough: the above equations refer
only to average values, while we want the motion to be approximately
Newtonian for single realizations of the stochastic process,
otherwise the model would not reproduce classical mechanics at the
macro-level. In Ref.~\cite{ab1} it has been proven, for Gaussian
solutions and in the case of an isolated system, that the variance
of $\langle Q \rangle_{t}$ associated to the motion of the center of
mass evolves as follows:
\begin{equation}
{\bf V}[\langle Q \rangle_{t}]  \; \simeq \; \left\{
\begin{array}{ll}
(1.1 \times 10^{-31}\,t/\makebox{sec})\; \makebox{m}^2\;\; & \makebox{for a $1 g$ object}, \\
& \\ (1.8 \times 10^{-59}\,t/\makebox{sec})\; \makebox{m}^2\;\; &
\makebox{for the Earth},
\end{array} \right.
\end{equation}
for $t < 2.0 \times 10^{4}$ sec, while for longer times it increases
like $t^3$. We see that for a macro--object and for very long times
(much longer that the time during which a system can be kept
isolated) the fluctuations are so small that, for all practical
purposes, they can be safely neglected; this is how {\it classical
determinism} is recovered within our stochastic model. Note thus
that, contrary to the behavior of the reduction mechanism, which is
amplified when moving from the micro- to the macro-level, the
fluctuations associated to the motion of microscopic particles
interfere destructively with each other, in such a way that the
diffusion process associated to the center of mass of an
$N$-particle system is much weaker than that of the single
components.

The above results imply that the actual values of $\langle Q
\rangle_{t}$ (and also of $\langle P \rangle_{t}$) are practically
equivalent to their stochastic averages, which obey
Eqs.~(\ref{emim}); we than have that $\langle Q \rangle_{t}$ and
$\langle P \rangle_{t}$ practically evolve according to the
classical laws of motion (in the appropriate physical situations)
for most realizations of the stochastic process. Since, for very
localized states like those having a spread given by
Eq.~(\ref{aval1}), $\langle Q \rangle_{t}$ represents the spot where
the wave function is concentrated, we reach the following
conclusion: in the macroscopic regime, the wave function of a
macroscopic system behaves, for all practical purposes, like a
point--like particle moving deterministically according to Newton's
laws of motion.

\subsection{Measuring situation}

In a recent paper \cite{ab2} we have analyzed the evolution of the
wave function as predicted by Eq.~(\ref{nlemp}), when a
macroscopic system acting as a measuring device interacts with a
microscopic quantum system in such a way to measure one of its
properties. The paper contains a mathematical analysis of the
situation, and proves, also giving precise estimates, the
following results:
\begin{enumerate}
\item\label{item:1} whichever the initial state of the microscopic
system, throughout the entire measurement process the center of
mass of the pointer is always extremely well localized and moves
as expected, from the ready state position to its final position.
\item\label{item:2} the only possible outcomes correspond to those given
by standard quantum mechanics, with probability infinitesimally
close to~$1$;
\item\label{item:3} the probability of getting a certain outcome is given
by the {\it Born probability rule} within an exceedingly high degree
of approximation;
\item\label{item:4} after the measurement, the microscopic system is in a
state which practically coincides with an eigenstate of the
observable which has been measured, corresponding to the eigenvalue
which has been observed.
\end{enumerate}
This proves rigorously what was the first goal of the original GRW
model: to provide a consistent solution to the measurement problem
of Quantum Mechanics.

\subsection{Identical particles}

The GRW model, as well as the QMUPL model previously discussed,
refers to a non-relativistic system with an arbitrary number of
{\it distinguishable} particles; the model has been successfully
generalized to include also {\it identical} particles. The best
known example is the CSL model \cite{csl} which is based on the
following stochastic differential equation:
\begin{equation} \label{eq:csl}
d\psi_{t} = \left[ -\frac{i}{\hbar} \, H\, dt + \sqrt{\gamma}\int
d^3 x \left( N({\bf x}) - \langle N({\bf x}) \rangle_{t} \right)
dW_{t}({\bf x}) - \frac{\gamma}{2} \int d^3 x \left( N({\bf x}) -
\langle N({\bf x}) \rangle_{t} \right)^2 dt \right]\psi_{t},
\end{equation}
where the symbol $\langle N({\bf x}) \rangle_{t}$ denotes the
quantum average of the operator $N({\bf x})$, which is an {\it
average density number} operator defined as follows:
\begin{equation}
N({\bf x}) \; = \; \left( \frac{\alpha}{2\pi} \right)^{3/2}
\sum_{s} \int d^3 y\, e^{- \frac{\alpha}{2} ({\bf x} - {\bf y})^2}
\, a^{\dagger}(s, {\bf y})\, a(s, {\bf y}),
\end{equation}
where $a^{\dagger}(s, {\bf y})$ ($a(s, {\bf y})$) is the creation
(annihilation) operator of a particle of spin $s$ at position ${\bf
y}$ of space. Note that, instead of having different Wiener
processes attached to each particle as in Eq.~(\ref{nlemp}), which
make the particles follow different histories and thus be
distinguishable, we now have a continuum of independent Wiener
processes $W_{t}({\bf x})$ (one for each point of space) which are
not attached to any particular particle, but only to their (average)
number density; hence the evolution respects the symmetry or
anti-symmetry properties of the wave function. The constant $\gamma$
has been set equal to $\lambda (4\pi/\alpha)^{3/2}$, where $\lambda$
is given in~(\ref{eq:num}), in order for GRW and CSL to coincide for
one particle.

The CSL model has been widely studied in the literature, and we
refer the reader to \cite{rev1} for the details. See \cite{tum1}
for a discrete, GRW-like, reduction model for identical particles.

\section{Relativistic dynamical reduction models}

The great challenge of the dynamical reduction program is to
formulate a consistent model of spontaneous wave function collapse
for relativistic quantum field theories; many attempts have been
proposed so far, none of which is as satisfactory as the
non-relativistic GRW model.

The first attempt~\cite{cslr} aimed at making the CSL model
relativistically invariant by replacing Eq.~(\ref{eq:csl}) with a
Tomonaga-Schwinger equation of the type:
\begin{equation} \label{eq:ts}
\frac{\delta\psi(\sigma)}{\delta \sigma(x)} = \left[
-\frac{i}{\hbar} \, {\mathcal H}(x) + \sqrt{\gamma} \left( {\mathcal
L}(x) - \langle {\mathcal L}(x) \rangle \right) V(x) -
\frac{\gamma}{2} \left( {\mathcal L}(x) - \langle {\mathcal L}(x)
\rangle \right)^2 \right]\psi(\sigma),
\end{equation}
where now the wave function is defined on an arbitrary space-like
hypersurface $\sigma$ of space-time. The operator ${\mathcal
H}(x)$ is the Hamiltonian density of the system ($x$ now denotes a
point in space-time), and ${\mathcal L}(x)$ is a {\it local}
density of the fields, on whose eigenmanifolds one decides to
localize the wave function. The $c$-number function $V(x)$ is a
stochastic process on space-time with mean equal to zero, while
the correlation function---in order for the theory to be Lorentz
invariant in the appropriate stochastic sense~\cite{cslr}---must
be a Lorentz scalar. And here the problems arise!

The simplest Lorentz invariant choice for the correlation function
is:
\begin{equation} \label{eq:cf}
{\bf E}[V(x) V(y)] = \delta^{(4)}(x - y),
\end{equation}
which however is not physically acceptable as it causes an infinite
production of energy per unit time and unit volume. The reason is
that in Eq.~(\ref{eq:ts}) the fields are {\it locally} coupled to
the noise which, when it is assumed to be {\it white}, is too
violent, so to speak, and causes too many particles to come out of
the vacuum. To better understand the situation, let us go back to
the non-relativistic Eq.~(\ref{eq:csl}): also there we basically
have a white-noise process, which however is not coupled locally to
the quantum field $a^{\dagger}(s, {\bf y}) a(s, {\bf y})$, the
coupling being mediated by the smearing Gaussian function appearing
in the definition of $N({\bf x})$. One can compute the energy
increase due to the collapse mechanism, which turns out to be {\it
proportional} to $\alpha$. Now, if we want to have a local coupling
between the quantum field and the noise, we must set $\alpha
\rightarrow + \infty$, in which case the energy automatically
diverges to infinity.

The simplest way out one would think of, in order to cure this
problem of Eq.~(\ref{eq:ts}), is to replace the local coupling
between the noise and the quantum field by a non-local one, as in
the CLS equation~(\ref{eq:csl}); this procedure would essentially
amount to replacing the white noise field with a non-white one. In
both cases we need to find a Lorentz invariant function which either
smears out the coupling or replaces the Dirac-delta in the
definition of the correlation function~(\ref{eq:cf}). This however
is not a straightforward task, for the following reason.

One of the reasons why the third term $(\gamma/2) \left( {\mathcal
L}(x) - \langle {\mathcal L}(x) \rangle \right)^2$ appears in
Eq.~(\ref{eq:ts}) is to guarantee that the collapse mechanism occurs
with the correct quantum probabilities (for those experts in
stochastic processes, the third term is such that the equation
embodies an appropriate martingale structure); if we change the
noise, we then have to change also the third term, and it turns out
that we have to replace it with a {\it non-local} function of the
fields~\cite{rim, pea1}. But, having a non-local function of the
fields jeopardizes the entire (somehow formal) construction of the
theory based on the Tomanaga-Schwinger equation, as the
integrability conditions are not automatically satisfied, and it is
very likely that the model will turn out to be inconsistent.

What we have briefly described is the major obstacle to finding a
relativistic dynamical reduction model. We want to briefly mention
three research programs which try to overcome such an impasse.

P. Pearle has spent many years in trying to avoid the infinite
energy increase of relativistic spontaneous collapse models, e.g.
by considering a tachyonic noise in place of a white noise as the
agent of the collapse process~\cite{pea2}, obtaining suggestive
results. Unfortunately, as he has recently admitted~\cite{pea1},
the program so far did not succeed.

Dowker and Henson have proposed a spontaneous collapse model for a
quantum field theory defined on a 1+1 null lattice~\cite{dow1,
dow2}, studying issues like the non-locality of the model and the
no-faster-than-light constraint. More work needs to be done in
trying to apply it to more realistic field theories; in particular,
it would be important to understand if, in the continuum limit, one
can get rid of the divergences which plague the relativistic CSL
model.

In a recent paper~\cite{tum2}, generalizing a previous idea of
Bell~\cite{bell4}, Tumulka has proposed a discrete, GRW-like,
relativistic model, for a system of $N$ non-interacting particles,
based on the multi-time formalism with $N$ Dirac equations, one per
particle; the model fulfills all the necessary requirements, thus it
represents a promising step forward in the search for a relativistic
theory of dynamical reduction. Now it is important to understand
whether it can be generalized in order to include also interactions.

It is rather discomforting that, after so many years and so many
efforts, no satisfactory model of spontaneous wave function collapse
for relativistic quantum field theories exists. And some have
started to wonder whether there is some fundamental incompatibility
between the dynamical reduction program and relativity. In this
regard, we mention the analysis of Ref.~\cite{ghi1}, where a toy
model of spontaneous wave function collapse is analyzed: the
collapse mechanism is supposed to occur instantaneously along all
spacelike hypersurfaces crossing the center of the jump process; in
spite of this superluminal effect, the whole picture is perfectly
Lorentz invariant, it agrees with quantum mechanical predictions, it
does not lead to any contradiction, e.g. it does not allow
faster-than-light signalling and, moreover, different inertial
observers always agree on the outcomes of experiments.
Unfortunately, the missing piece (which would make the toy model a
real physical model) is the dynamics for the reduction mechanism; in
any case, this model suggests that there is no reason of principle
forbidding the relativistic reduction program.

\section{Open questions and future developments}

Apart the important issue of finding satisfactory relativistic
dynamical reduction models, there is still much work to be done
and many open questions to be answered, at different levels,
ranging from mathematical to experimental physics. We conclude
this paper with a (partial) list some of interesting open
problems.

\subsection{Open questions: Mathematical Physics}

Let us consider once more Eq.~({\ref{nlemp}), which for a single
particle reads:
\begin{equation} \label{nlemp1}
d\,\psi_{t}(x) \; = \; \left[ -\frac{i}{\hbar}\, H\, dt +
\sqrt{\lambda}\, ( q - \langle q \rangle_{t})\, d W_{t} -
\frac{\lambda}{2}\,  ( q - \langle q \rangle_{t})^2 dt \right]
\psi_{t}(x);
\end{equation}
this is the simplest known continuous generalization of the
original GRW model, and existence and uniqueness of solutions have
been proved already in a number of theorems \cite{gg,hol}; still,
many important properties have not been studied yet. For example:
is it possible to write down explicitly the general solution of
Eq.~(\ref{nlemp1}) for the three most significant, usually exactly
solvable, physical systems, namely the free particle ($H =
p^2/2m$), the harmonic oscillator ($H = p^2/2m + m \omega^2/2$)
and the hydrogen atom ($H = p^2/2m - e^2/r$)? What about the other
types of Hamiltonian operator which can be solved analytically in
the standard quantum case?

Another important type of problems concerns the large time behavior
of the solution of Eq.~(\ref{nlemp1}). In the {\it free-particle}
case, various authors in different ways proved or were close to
proving, with a different degree of rigor, that any initial state
belonging to the domain of the equation, with the possible exception
of a subset of measure zero, converges to a Gaussian wave function
with a fixed spread both in position and momentum, while the
respective mean values diffuse. Apart from the free particle, for
which class of Hamiltonian operators $H$ does almost any initial
state converge to some fixed state with a finite spread in position
and momentum? Does this class contain the physically most
significant Hamiltonian operators? With what rate does an initial
state converge to the asymptotic state? Does the amplification
mechanism work as expected, i.e. in such a way that the bigger the
particle, the faster the collapse?

\subsection{Open questions: Theoretical Physics}

I think that many researchers in this field consider the dynamical
reduction program as a first important step towards a formulation of
a new theory for microscopic physical processes, which supersedes
quantum mechanics; the big question is: what does this theory look
like? There has been a lot of speculation in this regard, which
dates back to Einstein who thought of quantum mechanics as a
provisional theory which eventually will end up in being a
statistical approximation of a deeper theory; as far as I know the
only concrete proposal along these lines is the one put forward by
S. Adler in his recent book~\cite{adb}. He assumes precisely that
quantum mechanics is not a fundamental theory of nature but an
emergent phenomenon arising from the statistical mechanics of matrix
models with a global unitary invariance. The book is entirely
devoted to showing how that idea can be implemented within a
concrete (and highly sophisticated) mathematical framework, and we
invite the reader to look at it for all necessary details.

Another important issue, related to the previous one, is the
following. In all dynamical reduction models so far developed, the
stochastic process responsible for the collapse of the wave
function is a sort of mathematical entity without an existence on
its own, whose only job is to localize the wave function. I find
it very tempting to imagine that this field is {\it real}, that it
has its own equations of motion and acts on the quantum system,
but also that the quantum system acts back on it. It is also
tempting to say that this field is not a new field of nature, but
the only field that has not been successfully quantized yet, i.e.
the gravitational field. The research has already moved in this
direction~\cite{rev2,gra1,gra2,gra3}, and it is very exciting and
worthwhile pursuing it. It could also clarify a rather delicate
issue connected with the violation of the energy conservation
principle in dynamical reduction models.

As discussed in~\cite{ab1}, the collapse mechanism induces a sort of
diffusion process on the wave function in momentum space, which
makes it pick up higher and higher components in momentum, which in
turn show up as an increase of the energy of the system. With the
choice made in~(\ref{eq:num}) for the parameters $\lambda$ and
$\alpha$, such a violation is very tiny and hardly detectable with
present day technology; still, it is present and some people find it
disturbing. Now, if the stochastic field has to be regarded as a
real physical field with a reality on its own and its own equations
of motion, it seems natural to think that we are making a mistake in
assuming that the energy of a quantum system should be conserved; in
the calculations we should instead consider the energy of the
stochastic field {\it together} with the energy of the quantum
system, as the global energy which should be conserved. In this way,
there is serious hope to restore the principle of energy
conservation within dynamical reduction models. A first promising
step in this direction has been put forward in Ref.~\cite{biv}.

\subsection{Open questions: Phenomenology and Experiments}

\begin{table}
\caption{\label{tab2} The table shows the {\it upper bounds} on the
possible numerical value of the collapse parameter $\lambda_{0}$,
set by present-day observational data. The values are taken
from~\cite{adp}. E.g. for fullerene experiments, the number $5
\times 10^{12} \lambda_{0}$ means that the upper bound is $5 \times
10^{12}$ times larger than the standard value $\lambda_{0}$ as given
in Sec. 4.}
\begin{center}
\begin{tabular}{ccccc} \br
Fullerene & Decay of & Radiation by & 11 KeV photons & Proton \\
diffraction & super-currents  & free electrons & from Ge & decay\\
\mr $5 \times 10^{12} \lambda_{0}$  & $10^{14} \lambda_{0}$ &
$10^{12}
\lambda_{0}$ & $3 \times 10^{14} \lambda_{0}$ & $10^{18} \lambda_{0}$\\
\br Hydrogen & Heating of & I.G.M. & Inter-stellar &\\
dissociation & protons & & dust grains &\\
\mr $4 \times 10^{17} \lambda_{0}$ & $10^{12} \lambda_{0}$ &
$10^{8 \pm 1} \lambda_{0}$ & $10^{15} \lambda_{0}$ & \\ \br
\end{tabular}
\end{center}
\end{table}

Dynamical reduction models, by modifying the Schr\"odinger equation,
are predictively different from standard quantum mechanics; it
becomes then interesting to look for situations where it would be
easier to test these models against the standard quantum theory.
Indeed, the importance of such a research goes far beyond the
dynamical reduction program itself, as it ultimately would aim at
testing one of the most characteristic traits of quantum mechanics,
namely the superposition principle.

In a recent paper~\cite{adp}, S. Adler has done an exhaustive and
up-to-date review of the most plausible scenarios where it is more
likely to detect possible violations of the superposition
principle, and the presence of a spontaneous collapse mechanism.
These results are summarized in table~\ref{tab2}.

Basically, there are two scenarios where it is more convenient to
look for possible GRW-effects: high precision experiments on
micro-systems, and cosmological data; the first are reported on the
upper row, and the second on the lower row. As we see, for the
standard values given in Eq.~(\ref{eq:num}), the constraints are
rather weak and there is no hope that in the near future such
effects can be possibly tested.

However, in~\cite{adp} Adler notices that, given~(\ref{eq:num}), a
wave function is not reduced when a latent image is formed, in
photography or etched track detection. Since one would think it very
natural to assume the localization process to occur already at the
stage of latent image formation (as a latent image can be safely
stored for very long times and only afterwards developed or etched),
Adler suggests to increase the standard numerical value of
$\lambda_{0}$ by order of $2 \times 10^{9 \pm 2}$, to guarantee the
collapse to occur already at this stage. This is a very suggestive
hypothesis, since it implies that in the very near future technology
will be available, which will allow for a test of dynamical
reduction models. \\ \noindent
{\bf Acknowledgements.} The author wishes to thank the organizers
for having been invited to the {\it Third International Workshop
DICE2006}, for the pleasant atmosphere, and for the efforts put in
organizing it. Many thanks also to S.L. Adler, L. Diosi, D. D\"urr,
G.C. Ghirardi and R. Tumulka, for many stimulating discussions.

\end{document}